\documentclass{article}

\usepackage{arxiv}

\usepackage[utf8]{inputenc} 
\usepackage[T1]{fontenc}    
\usepackage{hyperref}       
\usepackage{url}            
\usepackage{booktabs}       
\usepackage{amsfonts}       
\usepackage{microtype}      
\usepackage{amsmath}        
\usepackage{graphicx}

\title{Influence of the ectonic emission mechanism on the collisionless expansion of a multicomponent cathode plasma}

\author{
 Vasily Yu. Kozhevnikov \\
  Laboratory of Theoretical Physics\\
  Institute of High Current Electronics\\
  Tomsk, 634055 \\
  \texttt{Vasily.Y.Kozhevnikov@ieee.org}
}

\begin{document}
\maketitle
\begin{abstract}
In this paper, the authors delve into the collisionless kinetic theory of planar vacuum diodes to investigate how a pulse-periodic (ectonic) cathode plasma emission mechanism influences the non-thermal expansion of the cathode plasma and "anomalous" ion acceleration. Another aspect explored in this paper is how pulse-periodic emission affects the comparative dynamics of motion of multi-charged ions from the cathode material. Using kinetic theory, we conclude that the ectonic emission mechanism affects both the macroscopic picture of plasma expansion and the intricacies of electric potential redistribution over one complete emission period. 
\end{abstract}


\section{Introduction}
Over the last seventy years, in the field of vacuum discharge physics, the initial two phases, namely breakdown and spark, have garnered significant interest, particularly within pulsed power technology. Understanding breakdown is crucial for enhancing the electrical insulation of various devices such as pulse generators, electron and ion accelerators, microwave devices, and pulsed X-ray generators. Developing compact and dependable pulsed power systems necessitates a thorough understanding of vacuum breakdown mechanisms. A widely debated topic in the transient phase of vacuum discharge is the phenomenon known as "anomalous ion acceleration". First identified by the experimental group led by Plutto in a plasma diode \cite{b1}, this phenomenon entails the abrupt emergence of positively charged ion bursts that undergo peculiar acceleration from the cathode towards the anode. While the electric potential distribution in vacuum diodes inherently accelerates the cathode plasma's electron component, the higher anode potential, from the standpoint of ion motion, represents a potential barrier. Consequently, the ion transport from cathode to anode is called "anomalous ion acceleration". During the vacuum breakdown phase, the typical plasma expansion velocities range from $1 \cdot 10^6$ to $6 \cdot 10^6$~cm/s, and the ions contribute up to 12~\% of the total charge transported to the anode. Their typical ion kinetic energies in diodes correspondingly span from tens to thousands of electron volts. Such ion flows with comparable high energies have been consistently observed in vacuum discharges. This effect has been reliably confirmed experimentally and is widely reported in the literature on vacuum discharges, e.g., \cite{b2, b3}.

Theoretical deliberations on anomalous ion acceleration pose a significant challenge. At present, theoretical hypotheses about the origin of ion flow directed towards the anode are broadly categorized into three classes: explosive, collisional, and electrodynamic \cite{b4}. From the standpoint of abstract theoretical constructs, all three perspectives can be considered as having grounds for existence. However, as was noted by academician G.A.~Mesyats \cite{b3}: "a common drawback of many proposed models is that the anomalous ion acceleration has been examined in isolation from the entirety of accompanying physical processes. Consequently, the conclusions are qualitative, and the initial assumptions are inconsistent with experimental data."

An important milestone in developing the theory of an expanding cathode plasma was the adoption of the most fundamental approach in plasma physics—physical kinetics. Such models were constructed for planar \cite{b5, b6} and geometrically non-uniform \cite{b7, b8} vacuum diodes with cathode plasma emission. They revealed that the cause of non-thermal expansion of the cathode plasma and anomalous ion acceleration is the emergence of a non-stationary, non-monotonic distribution of electric potential in the diode (a moving virtual cathode $\Delta \varphi < 0$). These studies elucidated the cause of anomalous ion acceleration and enabled the computation of average plasma expansion velocities. They determined the minor role of scattering collisions in this process.

Furthermore, the potential emergence of a "deep potential well" near the cathode $\left\lvert \Delta \varphi \right\rvert > U_0$ was demonstrated. The emission-center parameters entirely determine this effect and are independent of the applied voltage amplitude $U_0$ to the diode. Additionally, the presented theory elucidates why the hypothesis of a "potential hump" remains experimentally elusive \cite{b9} — the region of the accelerating potential's decay, $\Delta \varphi < 0$, is extremely narrow.

The findings from studies \cite{b5, b6, b7, b8} have proven crucial to understanding the mechanism underlying the initial stage of vacuum breakdown, particularly the expansion of the cathode plasma. However, these findings were all predicated on a significant assumption: that cathode plasma emission occurs continuously. Yet, experimental observations \cite{b3} have indicated that the electron current in explosive electron emission manifests in discrete bursts. These bursts of plasma emanating from explosive emission centers can be analogously conceptualized as emission quasi-particles, termed "ectons" \cite{b10}. The concept of ecton quasi-particles arises from observations of vacuum sparks and arcs, as processes at the emission center during explosive emission often display cyclical behavior \cite{b11}: plasma emission initiates, persists for a certain duration, halts, and then resumes in cycles. Electron-optical observations in chronographic mode further underscore the discontinuous nature of illumination of emission centers. For instance, illumination cycles on and off at least five times within several tens of nanoseconds, as reported in the cited paper \cite{b12}. For a copper cathode, illumination follows a 5 ns periodicity during the emission center's active lifetime of approximately 3-5~ns \cite{b13}. Pulse-periodic emission from the cathode induces intermittent emission current, causing a redistribution of electric potential at the cathode. As the initiation process leading to the formation of a mobile virtual cathode $\Delta \varphi < 0$ involves the appearance of a small negative space charge at the periphery of the cathode plasma cluster, the decline of emission current to zero can introduce significant adjustments to the overall cathode plasma expansion and anomalous ion acceleration dynamics.

This study explores the potential influence of the ectonic (pulse-periodic) plasma emission mechanism from the cathodic emission center on the cathode plasma expansion process and the ectonic mechanism's involvement in anomalous ion acceleration. From the standpoint of physical kinetics, the macroscopic picture of plasma expansion and the detailed dynamics of instantaneous electric potential values and particle distribution functions over a complete period of variation in the cathode emission are investigated. The proposed theory demonstrates that the previously identified collisionless electrostatic mechanism of plasma expansion remains predominant even when the emission region operates in a pulsating mode. Additionally, the study reveals that the periodic interruption characteristic of ectons' emission current does not significantly affect the anomalous acceleration of individual plasma ion components comprising ions of various charges.

\section{Theoretical model}
Let us consider the vacuum diode as a planar discharge gap formed by flat-parallel electrodes (cathode and anode). We will assume that the interelectrode gap is denoted by $D$, and the cross-sectional area of the diode is denoted by $S$. Then the distribution functions that characterize the physical states of the electron and ion ensembles will depend exclusively on the coordinate $x$, the longitudinal momentum component $p_x$, and the time variable $t$. For specificity, let us assume that the emission of a quasi-neutral multicomponent plasma from the cathode ($x = 0$) occurs at the initial time moment $t = 0$ and is characterized by the time-variable number density $n(t)$ and the temperatures $T_e$ and $T_i$ of electrons and ions, respectively.

Below, it is assumed that the cathodic emission plasma has a three-component composition, i.e., it includes electrons, singly charged ions, and doubly charged ions. Such a plasma charge composition is not abstract but is typical of cathodes made of magnesium, zinc, germanium, strontium, indium, and many other materials \cite{b14}. Here, the calculations employ data for zinc ($m_i = $65.38~a.m.u.), in which, according to experimental findings, 80 percent of the total ion fraction consists of singly charged ions $Zn^+$, while the remaining fraction comprises doubly charged ions $Zn^{++}$ \cite{b14}.

In this study, we adhere to the general methodology of physical kinetics of processes in plasma, wherein the components of the cathode plasma are characterized by corresponding single-particle distribution functions governed by collisionless Vlasov equations \cite{b15}. We assume the vacuum diode is in the absence of an external magnetic field. Thus, the self-consistent treatment of electromagnetic particle interactions is reduced to the mutual influence of electrostatic forces between particles, which is incorporated through the consistent inclusion of the Poisson equation into the system of equations of the mathematical model

\begin{equation}
\begin{cases}
\frac{\partial f_e}{\partial t} + \frac{p_x}{m_e} \frac{\partial f_e}{\partial x} - qE_x \frac{\partial f_e}{\partial p_x} = 0, \\[6pt]
\frac{\partial f_i^+}{\partial t} + \frac{p_x}{m_i} \frac{\partial f_i^+}{\partial x} + qE_x \frac{\partial f_i^+}{\partial p_x} = 0, \\[6pt]
\frac{\partial f_i^{++}}{\partial t} + \frac{p_x}{m_i} \frac{\partial f_i^{++}}{\partial x} + 2qE_x \frac{\partial f_i^{++}}{\partial p_x} = 0, \\[6pt]
\frac{\partial^2 \varphi}{\partial x^2} = -\frac{q}{\varepsilon_0} \left( n_+ + 2n_{++} - n_e \right), \quad E_x = -\frac{\partial \varphi}{\partial x},
\end{cases}
\label{eq:VP_system}
\end{equation}

where $f_e$ is the electron distribution function (EDF), $f_i^+$ and $f_i^{++}$ are the single and double charged ion distribution function (IDFs), $q$ is the electron charge, $E_x$ is the electric field, $\varphi$ is the electrostatic potential, $\varepsilon_0$ is the vacuum dielectric permittivity, $m_{e, i}$ are the rest masses of the electron and ion, respectively, and $n_{e,+,++}$ are electrons and ions number densities, respectively, which are determined as the corresponding distribution functions zero-moments

\begin{equation*}
n_{e,i}(x,t) = \int_{-\infty}^{\infty} f_{e,i}(x,p_x,t) \, dp_x.
\end{equation*}

The vacuum diode is considered to be connected to a voltage source $U(t)$ through a ballast resistance $R$. At time point $t = 0$, the voltage source produces a positive electric potential $U(t)$ with an amplitude $U_0$ and a short front edge of 0.1~ns. Under these conditions, the solution of a Poisson equation in the referenced planar geometry can be expressed in quadratures \cite{b6}

\begin{equation}
\begin{aligned}
E_x(x, t) = -\frac{U(t) - j(t)SR}{D} &+ \frac{q}{\varepsilon_0 D} \int_{0}^{D} \int_{0}^{x} \left[ n_e(x', t) - n_+(x', t) - 2n_{++}(x', t) \right] dx' \, dx \\
&- \frac{q}{\varepsilon_0} \int_{0}^{x} \left[ n_e(x', t) - n_+(x', t) - 2n_{++}(x', t) \right] dx', \\[8pt]
\varphi(x, t) = \frac{x}{D} \bigl( U(t) - j(t)SR \bigr) &- \frac{x}{D} \cdot \frac{q}{\varepsilon_0} \int_{0}^{D} \int_{0}^{x} \left[ n_e(x', t) - n_+(x', t) - 2n_{++}(x', t) \right] dx' \, dx \\
&+ \frac{q}{\varepsilon_0} \int_{0}^{x} \int_{0}^{x'} \left[ n_e(x'', t) - n_+(x'', t) - 2n_{++}(x'', t) \right] dx'' \, dx'.
\end{aligned}
\label{eq:Poisson_solution}
\end{equation}

where $j(t)$ represents the total current density in the electrical resistive circuit, which is determined as the superposition of the convective current densities of electrons and ions computed as the following EDF/IDF moments

\begin{align*}
j_e(x, t) &= q \int_{-\infty}^{\infty} \frac{p_x}{m_e} f_e(x, p_x, t) \, dp_x, \\
j_+(x, t) &= q \int_{-\infty}^{\infty} \frac{p_x}{m_i} f_i^+(x, p_x, t) \, dp_x, \\
j_{++}(x, t) &= 2q \int_{-\infty}^{\infty} \frac{p_x}{m_i} f_i^{++}(x, p_x, t) \, dp_x,
\end{align*}

and the displacement current density obeying the current balance equation

\begin{equation*}
\varepsilon_0 \frac{\partial E_x}{\partial t} + j_+ + j_{++} - j_e = j(t)
\end{equation*}

According to the general methodology of the previous theoretical paper \cite{b6}, the Vlasov-Poisson system of equations' starting parameters are established to model the entry of cathode plasma from the emission boundary $x = 0$ (cathode) into the empty vacuum gap

\begin{equation}
\begin{gathered}
f_{e,i}(x, p_x, t = 0) = 0, \\[6pt]
f_{e,i}(x = 0, p_x, t) = \frac{n_0 \kappa_{e,i} \chi(t)}{\sqrt{2\pi m_{e,i} \, T_{e,i}}} e^{-\frac{p_x^2}{2m_{e,i}T_{e,i}}},
\end{gathered}
\label{eq:BCs}
\end{equation}

where $\kappa_{e, i}$ are the particle fraction constants defined from a quasi-neutrality condition and the ion fraction distribution, $\chi(t)$ is the function of an impulse-periodic signal with unit amplitude, characterizing the emission in the pulsed regime depicted below in Figure~\ref{fig:Figure_1}

\begin{figure}[ht]
\centering
\includegraphics[scale=0.4]{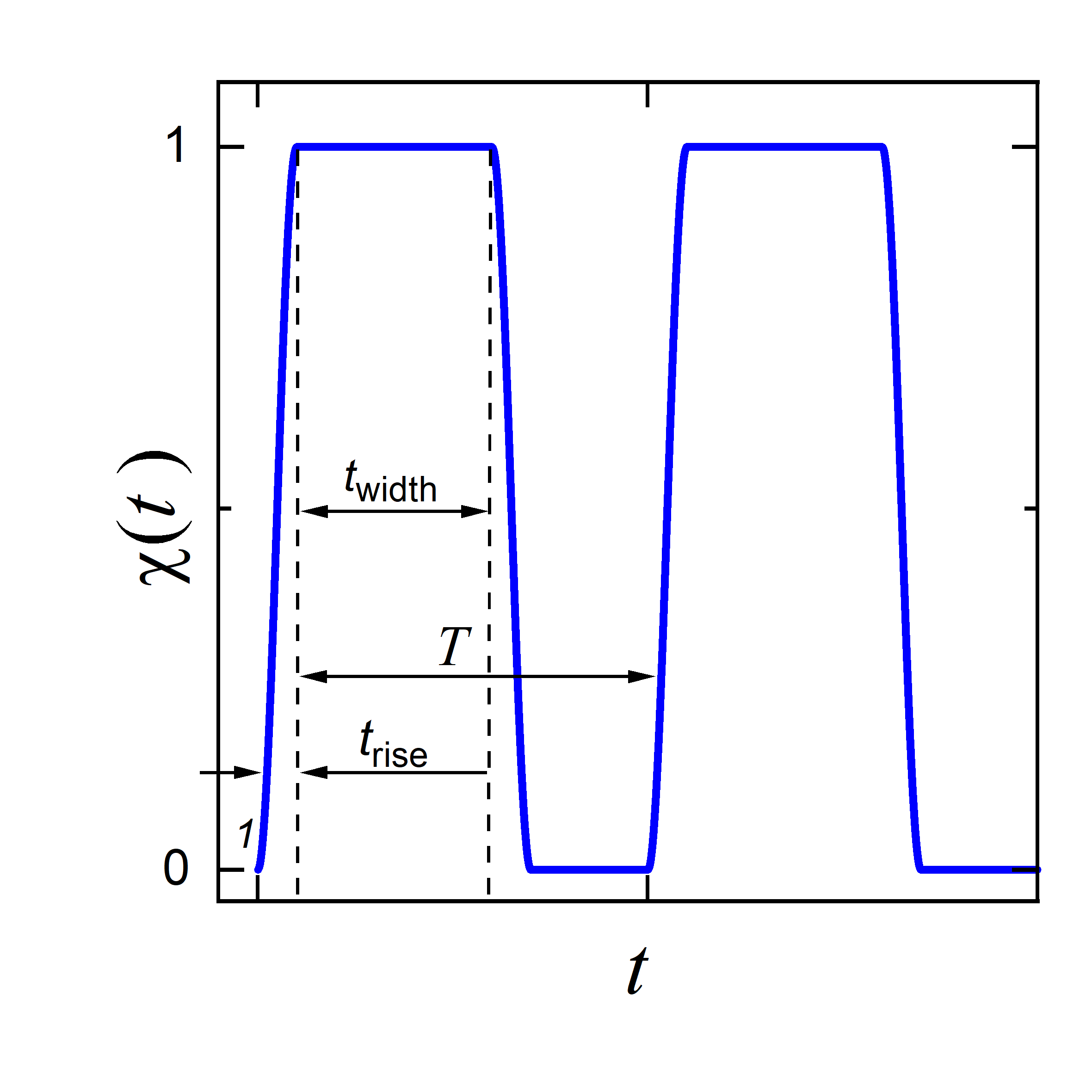}
\caption{\label{fig:Figure_1} A schematic representation of the non-stationary dependence of the dimensionless cathode plasma number density fluctuation parameter in the pulse-periodic emission mode (ectonic emission regime).}
\end{figure}

In the proposed model, the birth and demise of ectons are simulated by cyclic boundary conditions \eqref{eq:BCs}. We rely on phenomenological notions that the emerging explosive emission exhibits a cyclical nature: following the cathode microvolume explosion, and coupled with the heating of the emission zone by Joule heat, its cooling occurs via heat conduction, atomic evaporation, and emission cooling. This results in a reduction in the cathode and electric field, leading to cessation of emission. Summarising the main body of current experimental findings \cite{b3} regarding the formation of cathode emission in the form of so-called "ectons", without loss of generality, the average parameters of the function $\chi(t)$ are considered as $t_{rise} = 0.1$~ns, $t_{width} = 3$~ns, $T = 5$~ns.

\section{Numerical computation techniques}
The numerical solution of system \eqref{eq:VP_system}-\eqref{eq:BCs} follows standard approaches commonly employed in the literature. Specifically, the set of Vlasov equations \eqref{eq:VP_system} is discretized and solved on a quasi-uniform grid in the phase-space variables
\[
x(\xi) = \frac{D}{2} + \frac{c\xi}{\sqrt{1 + d\xi^2}}, \quad c = \frac{D}{2\xi_*} \left( \frac{D}{2} - a \right) \sqrt{\frac{1 - \xi_*^2}{a(D - a)}},
\]

\[
d = \frac{ \left( \frac{1}{2}D - a \right)^2 - \frac{1}{4}D^2\xi_*^2}{a (D-a) \xi_*^2}, \quad \xi \in [-1, 1],
\]

where $\xi$ is a parameter of a homogeneous grid in the interval $[-1, 1]$, $\xi_*$ is a numerical parameter determining the proportion of grid elements in the near-electrode layers, and $a$ is the characteristic size of the near-electrode computational layer. In the considered problem, for the interval of $D = 1$~cm, parameter values of $a = 1$~mm and $\xi_* = 0.25$ were used based on the requirement that the minimum size of the computational grid near the cathode is 10-15~times smaller than the characteristic Debye length. We use a decomposition method (semi-Lagrangian numerical scheme) that iteratively splits each Vlasov equation into a set of shift equations in phase coordinates \cite{b16}. To improve time resolution, the 4th- and 6th-order accuracy-splitting methods have been used in the time-stepping procedure \cite{b17}. The paper presents numerical results from implementing OpenMP-parallel code in C. This approach was adopted due to the substantial computational demands of solving the Vlasov system of equations.

The computation results below are given for the vacuum diode $D = 1$~mm, $U_0 = 2$~kV with $n_0 = 10^{22}$ m$^{-3}$ cathode plasma number density. In these calculations, a quasi-homogeneous grid for phase-space coordinates $(x, p_x)$ of 2500 by 2000 cells was employed for electrons, and a grid of the same dimensions was used for both types of ions. The characteristic time step of a semi-Lagrangian numerical scheme was systematically altered within the range of 0.1 to 1~picosecond.

\section{Cathode plasma expansion in ectonic emission regime: an overall view}

\begin{figure}[p]
\centering
\includegraphics[scale=0.28]{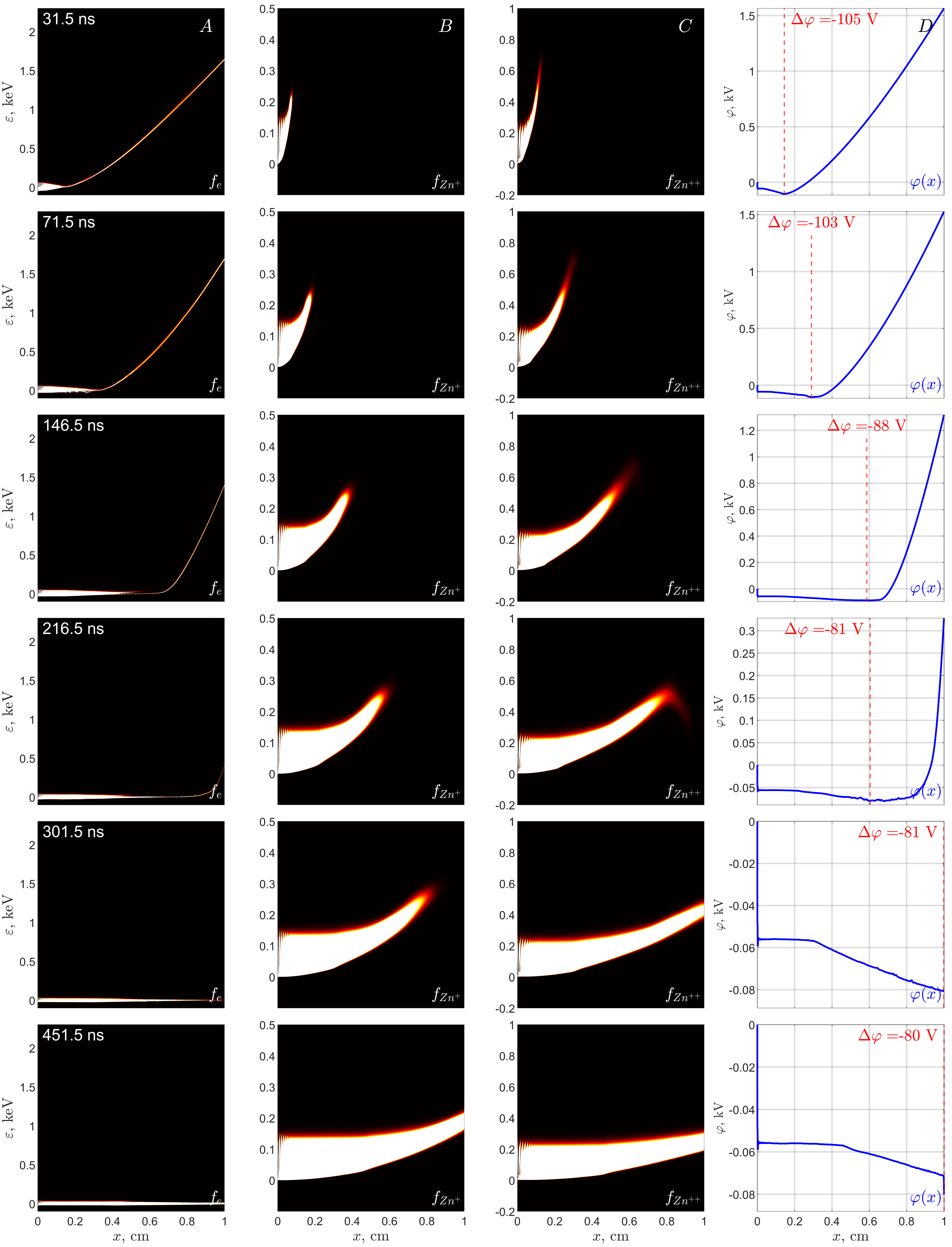}
\caption{\label{fig:Figure_2} Instant density plots: \textit{A}) - EDF, \textit{B}) – $Zn^+$ IDF, \textit{C}) – $Zn^{++}$ IDF, and curve 1 at \textit{D}) - spatial distributions of the electric potential at the selected time points. The red dashed line in plot \textit{D} indicates the position of the electric potential minimum ("virtual cathode"). EDF's brightness of color gamma levels is normalized to the maximum value at the anode, and the color scales for IDFs are restricted to an absolute value of $10^{39}$.}
\end{figure}

Figure~\ref{fig:Figure_2} illustrates the non-stationary density plots of electron and ion distribution functions and the instantaneous distribution of electric potential for the diode specified in the power supply circuit with a ballast resistance $R = 200$~ohms. This figure demonstrates the dynamic development of vacuum breakdown. The overall physical phenomenon that leads to plasma expansion and ion acceleration from the cathode to the anode in pulse-periodic emission is analogous to that in continuous emission. Specifically, at the periphery of the initially quasi-neutral cluster of a cathode plasma, quasi-neutrality is disrupted as the electron fraction is substantially lighter and thermalized ($T_e > T_i$). Local disruption of quasi-neutrality forms a region of negative space charge and, consequently, a region of negative electric potential $\Delta \varphi < 0$ ("virtual cathode"). The region of declining electric potential generates an electric force on ions, thereby initiating their accelerated movement toward the anode. Plasma fills the physical and virtual cathode region, effectively screening the near-cathode region. As the voltage source operates continuously, the electron current transport causes the virtual cathode region to move, further accelerating ions under the influence of a constant accelerating force; consequently, the plasma emission boundary shifts toward the anode. As can be observed, the mechanism of electrodynamic collisionless expansion of the cathode remains unchanged regardless of the assumption regarding the nature of emission.

The discussion regarding the average expansion velocity of the cathode plasma stems from experimental observations of the visual movement of the cathode plasma glow region within the vacuum gap. Theoretically, the cathode plasma is generally non-quasi-neutral because its components move at different speeds. In the context of the theoretical findings, a comparison can be made between the average velocity of the $Zn^{++}$ component and the experimental value obtained for vacuum breakdown in a diode with a zinc cathode. Here, a good agreement is observed, with the average velocity being approximately $1.1\cdot10^6$~cm/s \cite{b3, b18}, as derived from the calculation of the root-mean-square velocity of the $Zn^{++}$ component defined as follows $\overline{\nu}_{Zn^{++}} = \sqrt{ \int_{-\infty}^{\infty} v^2 f_i^{++} \, dv \Big / \int_{-\infty}^{\infty} f_i^{++} \, dv}$. An important observation is that the average velocities of the Zn+ and Zn++ ion components significantly exceed not only the thermal values of 1-5~eV but also the virtual cathode amplitude (multiplied by the electron charge), i.e., 100~eV in this case. This is due to the movement of the plasma emission boundary in an electric field of the "running wave" type, which is a key feature of the investigated plasma expansion mechanism and has also been noted in previous studies, where it was assumed that cathode emission takes on a continuous form \cite{b6, b7, b8}.

\begin{figure}[h]
\centering
\includegraphics[scale=0.4]{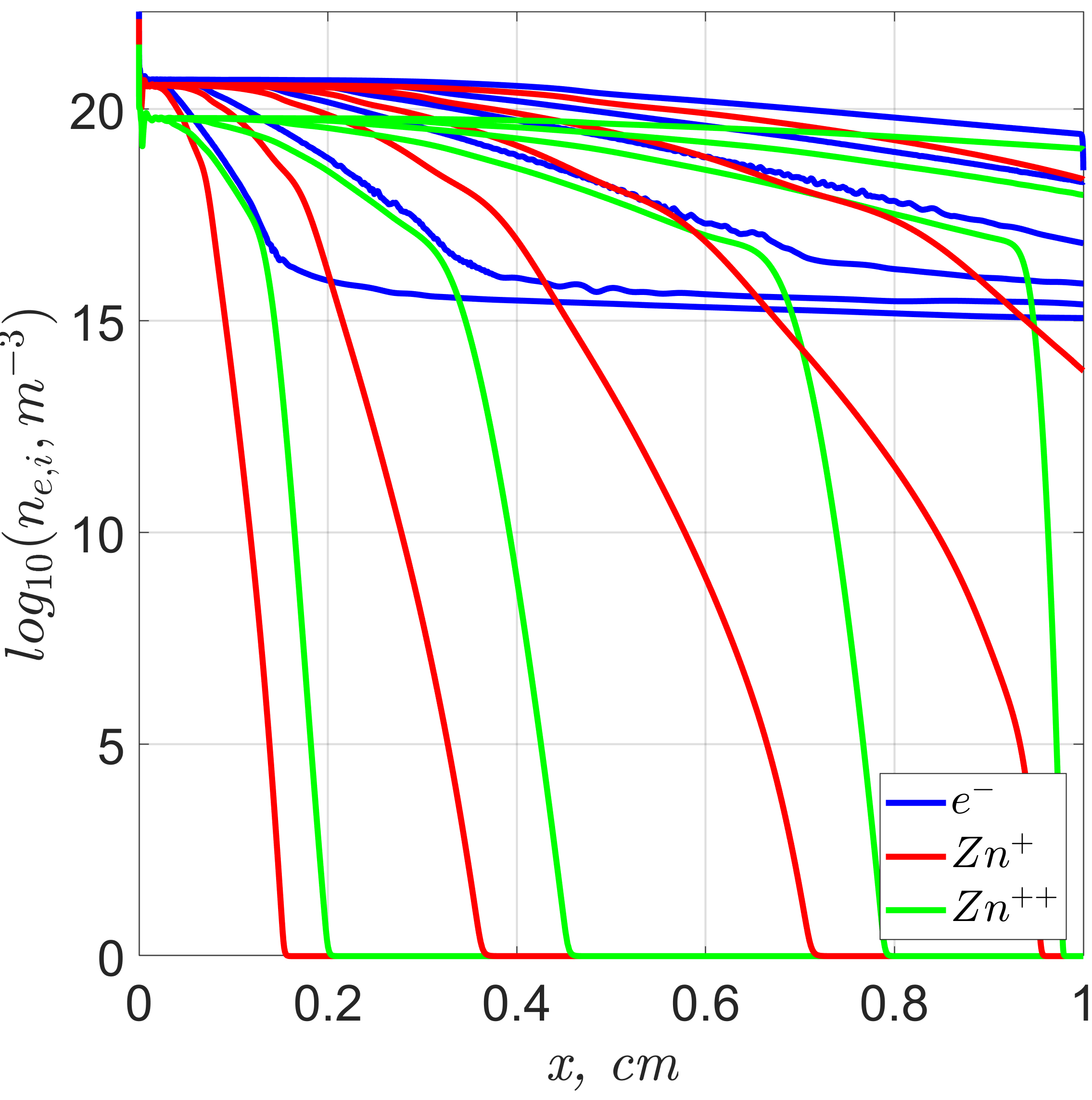}
\caption{\label{fig:Figure_3} Plasma component number densities given at the time points $t =$ 31.5, 71.5, 146.5, 216.5, 301.5, and 451.5~ns. Plots are given in decimal log scale (inverse hyperbolic sine function $\log_{10}(n_{e, i} + \sqrt{1+n^2_{e, i}})$ is used to avoid the $\log(-1)$ issue.}
\end{figure}

As previously stated, during the breakdown formation stage of the vacuum gap, the plasma filling it is not quasi-neutral. The expansion of the cathode plasma occurs at high (non-thermal) velocities, implying differences in the directed velocities of individual ion streams, as characterized by the ion's electric charge. Depending on the hypothesis regarding the physical mechanism of plasma expansion, it has been assumed that ions of different charges may move with approximately equal velocities (explosive hypothesis) or with velocities that differ significantly in magnitude (electrodynamic hypothesis). From an experimental standpoint, there is no unified perspective on this issue: the existing methodology and the experimental data need to be more consistent. For example, Davis and Miller \cite{b19} found that ion speed increases linearly with ion charge. However, in other studies (including experiments \cite{b20}), the measured ion speeds under specific conditions were essentially unaffected by the electric charge. The results from the modeling presented in this research support the viewpoint that the average speed of ions is directly related to the increase in charge Figure~\ref{fig:Figure_3}). Single-charged ions move at half the speed of doubly charged ions. Refinement of the emission mechanism does not significantly alter the dynamics of multi-charged ions, characterizing the overall dynamics of the collisionless expansion of the cathode plasma, as noted in \cite{b5}. In all the experiments listed, ion characteristics were measured based on data from ions behind the collector (anode).

As evident from Figure~\ref{fig:Figure_3}, $Zn^{++}$ ions reach the anode much earlier than $Zn^+$ ions. In the scenario depicted in Figure~\ref{fig:Figure_3}, this occurs around the time point of $t = 217$~ns when the anode potential value exceeds the average kinetic energy value of the arriving ion component (relative to the electron charge), i.e., $\varepsilon_i < qU(t)$. As a result, the $Zn^{++}$ ion flux experiences deceleration at the outer emission boundary of the expanding plasma while the $Zn^+$ ion flux continues to accelerate towards the anode. If, initially, when the plasma has not yet filled most of the gap, ion components move with average velocities that are linearly dependent on charge, then near the anode, higher-charge ions may experience earlier deceleration. A situation may arise in which the deceleration of the most accelerated (multi-charged) components aligns the velocities of ion streams with different charge compositions behind the collector. Another scenario is possible where the decline in the anode potential precedes the approach of highly charged ions. At the same time, they do not experience deceleration, resulting in significant separation of ions of different charges by velocity behind the collector. Thus, the ambiguity in experimental understanding of the dependence of the average velocity of multi-charged plasma components on the magnitude of electric charge does not indicate an error but rather reflects different experimental conditions, particularly differences in the electro-physical connection schemes of the vacuum diode. In the simplest case, this concerns selecting the value of the ballast resistance $R$ in the anode circuit.

By comparing the cathode plasma dynamics depicted in Figure~\ref{fig:Figure_2} and Figure~\ref{fig:Figure_3}, it can be inferred that after the plasma bridges the gap, the distribution of electric potential within the gap corresponds to quasi-neutrality, except for a narrow region near the anode where the potential drop is located, close to the amplitude value of the virtual cathode. 

From a macroscopic perspective, the relative dynamics of individual components of the cathode plasma in pulsed-periodic emission mode (ecton regime) and in continuous mode do not differ significantly \cite{b5}. The ecton emission type slightly reduces the average plasma expansion velocity (the velocity of the emission boundary movement), bringing it closer to the experimental values \cite{b18}. Of greater interest is the detailed plasma dynamics during the interval between ecton emissions, i.e., within one period $T$.

\begin{figure}[p]
\centering
\includegraphics[scale=0.46]{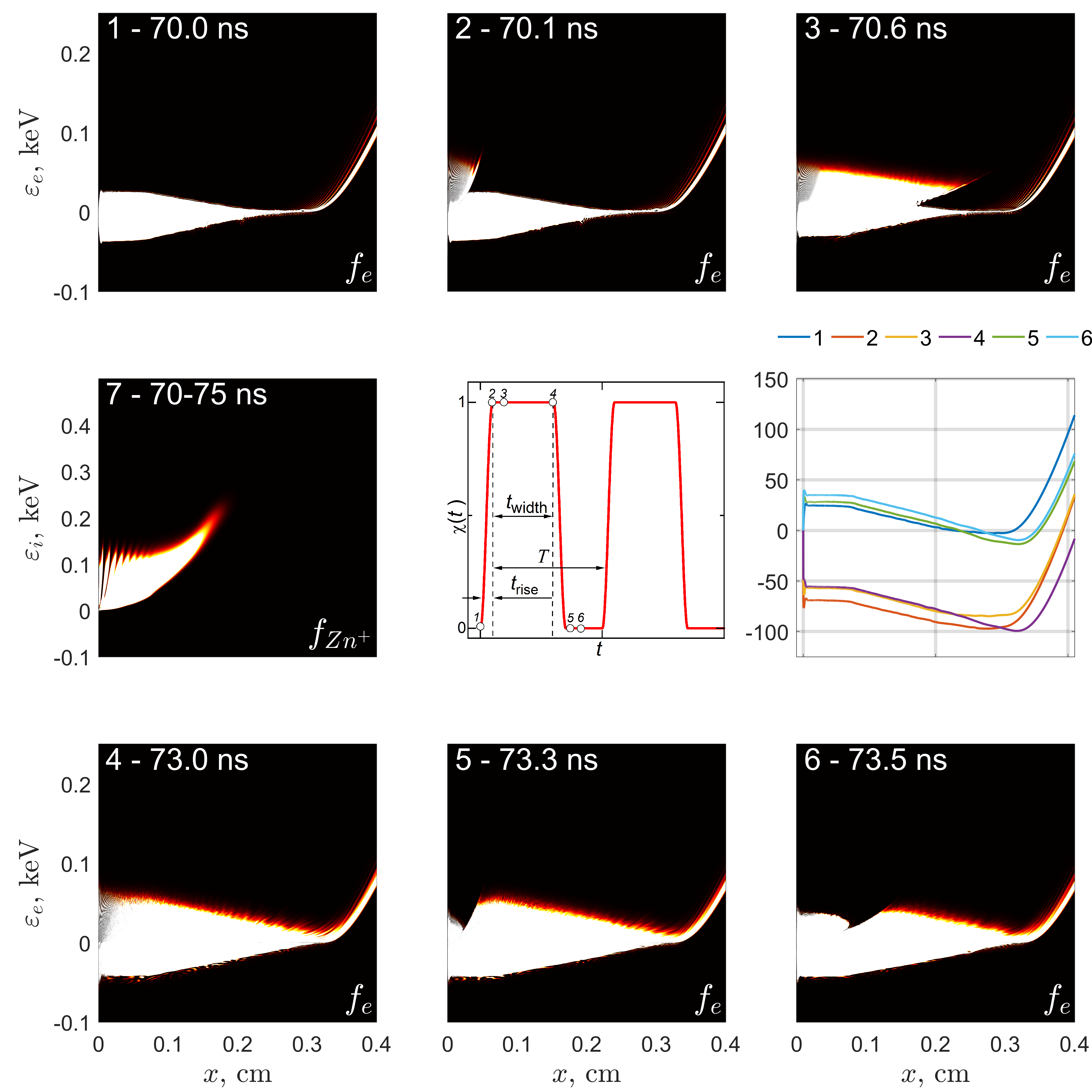}
\caption{\label{fig:Figure_4} The dynamics of the EDF, IDF, and electric potential changes within the cathode region at various time points over a single complete plasma emission pulse $T = 70-75$~ns from the cathode. The EDF graphs 1-6 correspond to the time points marked on the emission function diagram $\chi(t)$ and the electric potential distribution graph. The IDF for the specified time interval is depicted in Graph 7. The IDF color gamma is restricted to an absolute value of $10^{40}$.}
\end{figure}

\section{Cathode plasma expansion in ectonic emission regime: detailed view}

In this section, we endeavored to address the question: what will occur during characteristic timescales equivalent to one full emission period from the cathode $T$, which, according to experimental conceptions of ectons \cite{b18} in a vacuum discharge, is assumed to fit into a fixed range of $T = 1-10$~ns? Additionally of interest is the question of what precisely occurs during the so-called "dead time", a brief interval between emission bursts when the emission current from the cathode is absent. An additional computation was conducted to elucidate these inquiries in great detail for the EDF and the electric potential over the time interval $t = 70-75$~ns, encompassing roughly one cathode emission pulse. To mitigate the influence of an instantaneous redistribution of voltage drop between the diode and the ballast resistor, the calculation assumed that the voltage source (anode potential) is "ideal", i.e., $R = 0$. Other parameters of the vacuum diode, cathode plasma composition and its number density, voltage pulse parameters, and the pulse-periodic function $\chi(t)$ describing cathode emission were selected as analogous to those used previously (in Figure~\ref{fig:Figure_1}).

The results of these calculations are presented in Figure~\ref{fig:Figure_4}, where the EDF, IDF, and electric potential distribution are scaled near the cathode. Figure~\ref{fig:Figure_4} illustrates an example of what occurs during each emission period from the cathode, at $t = 70$~ns, corresponding to the onset of the next emission pulse. The emission current from the cathode begins to rise. Before this moment, emission was absent for around 2~ns, and the electric potential distribution exhibited characteristics typical of current transport from the quasi-neutral plasma layer, whose boundary is near the point of the electrostatic potential minimum ($\Delta \varphi = 0.5$~V) at $x = 0.3$~cm. From $t = 70.1$~ns, when the emission current begins to flow into the interval, until the end of the emission time at $t = 73$~ns, the potential distribution profile undergoes significant changes. At the presumed emission plasma boundary point, a virtual cathode is formed with an amplitude of $\Delta \varphi \approx - 100$~V (similar to those depicted in calculations in Figure~\ref{fig:Figure_2}), which exhibits all the characteristic features mentioned earlier: a sharp potential drop near the cathode of an approximately -100~V and a more gradual segment of the potential curve leading to the point of minimum where $\Delta \varphi \approx - 100$~V. During this time interval, fluctuations in the space charge of particles are observed (see the EDF density plot in Figure~\ref{fig:Figure_4}), which are caused by the birth of new ecton and the influx of an emission current into the quasi-neutral plasma, which is undergoing electron current transport conditions. Within a short interval of $t_{width} = 3$~ns, plasma expansion occurs due to both the accelerating electric force and the inertial properties of the ion component.

Figure~\ref{fig:Figure_4} depicts characteristic EDFs at the most significant time points throughout the $\chi(t)$ function. They exhibit distinctive, nontrivial features characterized by rapid changes in EDF that originate near the emission zone (cathode) and propagate towards the anode. Compared with continuous emission, the pulse-periodic nature of cathode plasma emission leads to oscillations in electron space charge, as evident in images 1-6 of Figure~\ref{fig:Figure_4}. These space charge oscillations cause significant variations in the electric potential in the gap between the cathode and the outer plasma emission boundary. During the 'dead time' of cathode emission $\chi(t)$, the virtual cathode essentially disappears in this region, i.e., $\Delta \varphi \approx 0$. Then, when emission resumes, the virtual cathode reappears at the outer emission boundary of the expanding plasma, regardless of its distance from the cathode. In other words, this behavior reflects the detailed plasma dynamics in the diode induced by the ecton mechanism: ecton decay leads to rapid 1-10~ns leveling of the local electric potential, and the birth of a new ecton recreates the virtual cathode with its previous amplitude, consequently leading to further plasma expansion. Notably, throughout this process, the anode voltage remains at a constant high value of 2~kV, ensuring continuous electron current transport. Moreover, the external field's continuous influence on electron transport smooths the virtual cathode dip during the pauses between electron emissions.

As mentioned earlier, the birth and decay of ectons result in local fluctuations of a negative space charge within the outer emission boundary of the cathode plasma. The characteristic lifetimes of ectons and the delays before the birth of a new quasi-particle are sufficiently short that over the emission period $T$, no significant global changes occur in the IDF distribution. This is because, within this time interval, ions tend to move towards the anode due to their inertial properties compared to electrons. The calculation in Figure~\ref{fig:Figure_2} vividly illustrates this phenomenon. However, fluctuations in the EDF in the vacuum diode also induce fluctuations in the IDF. Figure~\ref{fig:Figure_4} illustrates the distribution of IDF during the $t = 70-75$~ns period, which exhibits a characteristic serrated structure near the cathode. This feature arises from the multiple rapid redistributions of electric potential near the cathode during ecton-type emission. This feature is absent from studies describing continuous cathode emission \cite{b5, b6, b7, b8}.

\begin{figure}[h]
\centering
\includegraphics[scale=0.45]{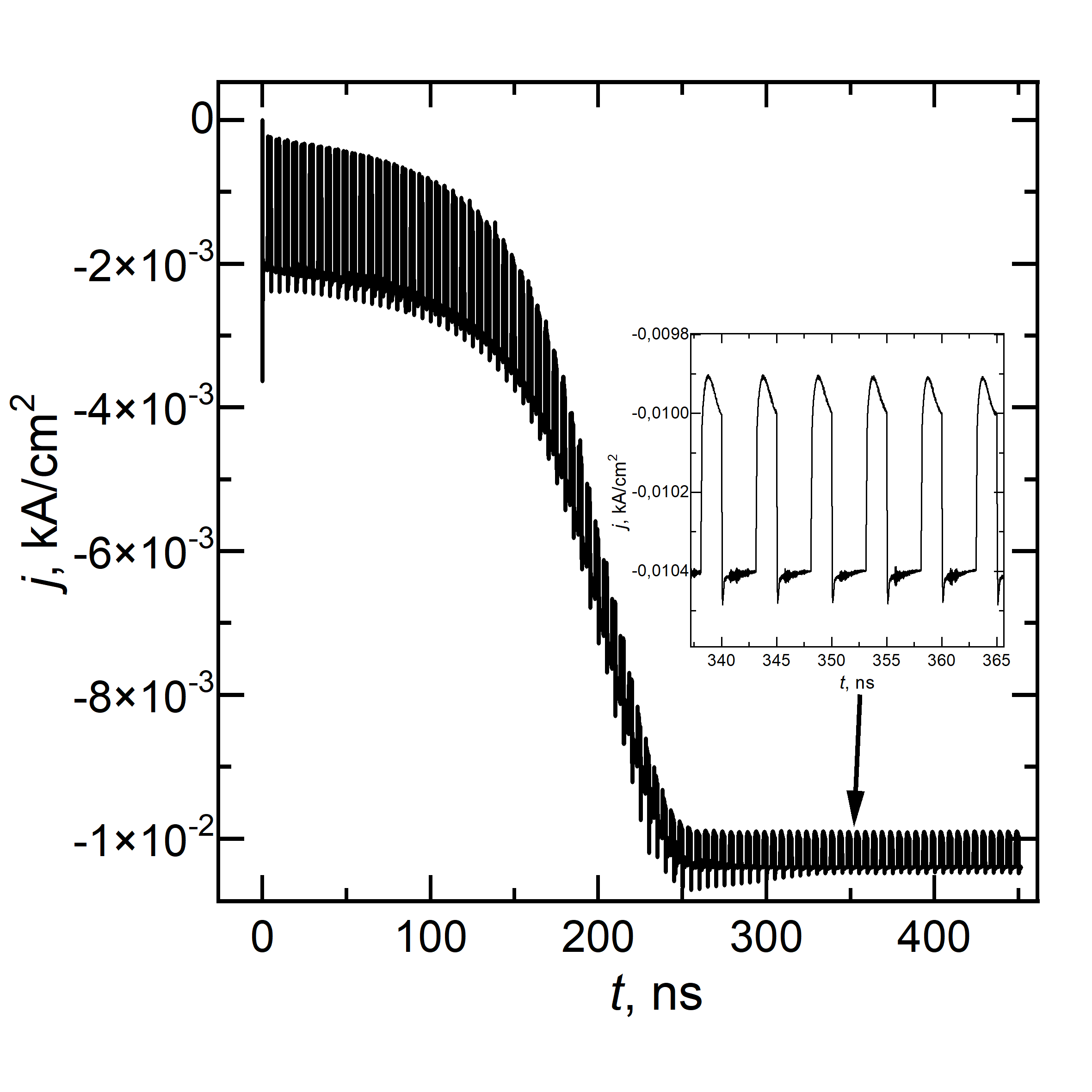}
\caption{\label{fig:Figure_5} Time-dependent full current density profile in vacuum diode with pulse-periodic emission from the cathode.}
\end{figure}

In summary, the processes of space charge fluctuations induced by pulse-periodic ecton emission are reflected in the time-dependent current flow through the diode. Figure~\ref{fig:Figure_5} plots the total current density in the vacuum diode within the power supply circuit with a ballast resistance of $R = 200$~ohms. It can be observed that throughout the entire vacuum breakdown process, the total current density in the diode forms a curve with pronounced frequency modulation. The modulation amplitude (Figure~\ref{fig:Figure_5}) is small (<~5~\%). Yet, its presence indicates the contribution of the ecton emission mechanism to the current flow, as the space charge oscillations have a period $T$ equal to the modulation period and the period of the $\chi(t)$ function.

\section{Conclusions}
When comparing the results of numerical modeling of vacuum gap breakdown phenomena in planar diodes with a continuous cathode emission mechanism \cite{b5, b6} and similar modeling of breakdown with an ecton emission mechanism, it was convincingly demonstrated that in both cases, the expansion mechanism of the cathode plasma is a collisionless field-induced process. Without loss of generality, it can be asserted that vacuum breakdown initiates when the emission center emits a cluster of quasi-neutral plasma that acquires negative space charge in its periphery. This leads to a local violation of electroneutrality and the formation of a virtual cathode, whose presence allows ions to move rapidly towards the anode. The region where the virtual cathode localizes forms a conditional boundary for electron-beam emission, with the expanding plasma effectively shielding the physical cathode. Due to the external potential difference applied to the diode, the gradual accumulation of ions leads to anode-directed movement of the virtual cathode (the outer emission boundary of the plasma), which in turn further facilitates the subsequent acceleration of ions towards the anode and further expansion of the cathode plasma. 

Detailed kinetic modeling of vacuum breakdown in the simple one-dimensional configuration of a planar diode with an ecton mechanism revealed the following findings:
\begin{itemize}
\item The formation of the initial cathode fall of the electric potential is characterized by durations significantly shorter than the typical ecton nucleation times. Therefore, plasma expansion from the cathode jet occurs at the moment of the first ecton's appearance;
\item The demise of one ecton and the pause before the birth of the next ecton result in a temporary disappearance of the virtual cathode region ($\Delta \varphi \approx 0$), yet it does not affect the plasma expansion from the cathode to the anode; 
\item Plasma expansion during the pauses between bursts of emission current occurs at the outer emission boundary of the quasi-neutral expanding plasma. For ions, this expansion is driven by the inertia of the ion component, owing to its kinetic energy acquired during the moments of ecton existence (emission);
\item The birth of each subsequent ecton leads to the formation of a new virtual cathode at the outer emission boundary. It accelerates a new portion of the plasma's ion component along the falling section of the electrostatic potential.
\end{itemize}

\section*{Acknowledgments}
The author expresses their sincere gratitude to Prof.~Dr.~Vladislav~Igumnov, whose enthusiasm, unwavering belief in the viability of this research area, and personal proactive support provided the necessary impetus to keep this project going despite all the difficulties.

The present study is a continuation of the research carried out within the framework of the State Task of the Ministry of Science and Higher Education of the Russian Federation on themes FWRM‑2021‑0007 and FWRM‑2021‑0014. This paper reports the latest post‑project results, which we believe provide new insights into the mechanism of ectonic emission and its influence on cathode plasma expansion.

\bibliographystyle{unsrt}

\begin{thebibliography}{1}

\bibitem{b1}
A. A. Plyutto.
\newblock Experimental Study of the Ion Component of a Vacuum Spark Plasma.
\newblock {\em Zhurnal Eksperimental'noi i Teoreticheskoi Fiziki}, 39:1589, 1960.

\bibitem{b2}
G. A. Mesyats.
\newblock {\em Pulsed Power}.
\newblock Springer, Boston, MA, 2005.

\bibitem{b3}
G. A. Mesyats.
\newblock {\em Explosive Electron Emission}.
\newblock Fizmatlit, Moscow, 2011.

\bibitem{b4}
G. Yu. Yushkov, A. S. Bugaev, I. A. Krinberg, and E. M. Oks.
\newblock On a mechanism of ion acceleration in vacuum arc-discharge plasma.
\newblock {\em Doklady Physics}, 46(5):307--309, 2001.

\bibitem{b5}
V. Kozhevnikov, A. Kozyrev, A. Kokovin, and N. Semeniuk.
\newblock The electrodynamic mechanism of collisionless multicomponent plasma expansion in vacuum discharges: From estimates to kinetic theory.
\newblock {\em Energies}, 14(22):7608, 2021.

\bibitem{b6}
A. Kozyrev, V. Yu. Kozhevnikov, N. S. Semeniuk, and A. O. Kokovin.
\newblock Kinetic Theory of the Expansion of a Multicomponent Cathode Plasma in a Planar Vacuum Diode.
\newblock {\em Plasma Sources Science and Technology}, 32:105010, 2023.

\bibitem{b7}
V. Yu. Kozhevnikov, A. V. Kozyrev, A. O. Kokovin, and N. S. Semenyuk.
\newblock Kinetic Model of Vacuum Plasma Expansion in a Cylindrical Gap.
\newblock {\em Plasma Physics Reports}, 49:1350, 2023.

\bibitem{b8}
J. Yao, V. Kozhevnikov, V. Igumnov, Z. Chu, C. Yuan, and Z. Zhou.
\newblock The kinetic theory of cathode plasma expansion in a spatially non-uniform geometric configuration of a vacuum diode.
\newblock {\em Plasma Sources Science and Technology}, 33:035006, 2024.

\bibitem{b9}
A. Anders.
\newblock The Evolution of Ion Charge States in Cathodic Vacuum Arcs.
\newblock In {\em International Symposium on Discharges and Electrical Insulation in Vacuum (ISDEIV)}, 2014.

\bibitem{b10}
G. A. Mesyats.
\newblock Ectons and their role in plasma processes.
\newblock {\em Plasma Physics and Controlled Fusion}, 47:A109, 2005.

\bibitem{b11}
G. A. Mesyats.
\newblock Ectons and their Role in Electrical Discharges in Vacuum and Gases.
\newblock {\em Le Journal de Physique IV}, 07:C4-93--C4-112, 1997.

\bibitem{b12}
Y. Y. Urike.
\newblock In {\em Proceedings of the V International Symposium of Discharges and Electrical Insulation in Vacuum}, pages 111--114, Poznan, Poland, 1972.

\bibitem{b13}
J. D. Cross, B. Mazurek, and K. D. Srivastava.
\newblock Photographic Observations of Breakdown Mechanism in Vacuum.
\newblock {\em IEEE Transactions on Electrical Insulation}, 18:230, 1983.

\bibitem{b14}
A. Anders.
\newblock Ion charge state distributions of vacuum arc plasmas: The origin of species.
\newblock {\em Physical Review E}, 55:969, 1997.

\bibitem{b15}
A. A. Vlasov.
\newblock The vibrational properties of an electron gas.
\newblock {\em Journal of Experimental and Theoretical Physics}, 8:291, 1938.

\bibitem{b16}
C. Z. Cheng and G. Knorr.
\newblock The Integration of the Vlasov Equation in Configuration Space.
\newblock {\em Journal of Computational Physics}, 22:330, 1976.

\bibitem{b17}
H. Yoshida.
\newblock Construction of Higher Order Symplectic Integrators.
\newblock {\em Physics Letters A}, 150:262, 1990.

\bibitem{b18}
G. A. Mesyats.
\newblock Ecton or electron avalanche from metal.
\newblock {\em Physics-Uspekhi}, 38:567, 1995.

\bibitem{b19}
W. D. Davis and H. C. Miller.
\newblock Analysis of the Electrode Products Emitted by dc Arcs in a Vacuum Ambient.
\newblock {\em Journal of Applied Physics}, 40:2212, 1969.

\bibitem{b20}
K. Tsuruta, K. Sekiya, and G. Watanabe.
\newblock Velocities of copper and silver ions generated from an impulse vacuum arc.
\newblock {\em IEEE Transactions on Plasma Science}, 25:603, 1997.

\end{thebibliography}

\end{document}